\documentclass[aps,pre,superscriptaddress,amsmath,preprint,showpacs]{revtex4}
\usepackage{amsmath}
\usepackage[dvips]{graphicx}

\newcommand{\be}{\begin{equation}}
\newcommand{\ee}{\end{equation}}

\newcommand{\eqs}[2]{Eqs. (\ref{#1}) \& (\ref{#2})}

\newcommand{\eq}[1]{Eq. (\ref{#1})} 
 
\newcommand{\Fig}[1]{Fig. (\ref{#1})} 
 \newcommand{\eqa}{\begin{eqnarray}}
\newcommand{\eeq}{\end{eqnarray}}

\begin{document}
\title{Modulation of waves due to charge-exchange collisions in magnetized
partially ionized space plasma}

\author{Dastgeer Shaikh\footnote{\tt Email : dastgeer.shaikh@uah.edu} and
G. P. Zank}
\affiliation{Department of Physics and 
Center for Space Plasma and Aeronomy Research, 
The University of Alabama in Huntsville, 
Huntsville, Alabama 35899, USA}

\begin{abstract}
A nonlinear time dependent fluid simulation model is developed that
describes the evolution of magnetohydrodynamic waves in the presence
of collisional and charge exchange interactions of a partially ionized
plasma. The partially ionized plasma consists of electrons, ions and a
significant number of neutral atoms.  In our model, the electrons and
ions are described by a single fluid compressible magnetohydrodynamic
(MHD) model and are coupled self-consistently to the neutral gas,
described by the compressible hydrodynamic equations. Both the plasma
and neutral fluids are treated with different energy equations that
describe thermal energy exchange processes between them. Based on our
self-consistent model, we find that propagating Alfv\'enic and fast/slow
modes grow and damp alternately through a nonlinear modulation
process.  The modulation appears to be robust and survives strong
damping by the neutral component.

\end{abstract}

\received{Sep 3,  2010, Revised Sep 8, 2010}
\maketitle
\section{Introduction}

Alfv\'en, fast and slow mode waves are characteristics of
magnetohydrodynamic (MHD) plasma. While the electromagnetic
fluctuations in Alfv\'en waves propagate predominantly along an
ambient or guide constant magnetic field, the fast/slow modes
propagates isotropically.
Although the linear properties of these waves are fairly well
understood \cite[]{pks1,pks2,pks3}, their nonlinear evolution in a
complex partially ionized environment, as occurs often in
space/astrophysical plasmas, remains relatively uninvestigated.  In
partially ionized (consisting of ions, electrons and a significant
neutral gas presence) space and astrophysical plasmas, these waves can
interact with the neutral gas through processes such as charge
exchange, collisions, ionization and recombination etc. Hence the wave
properties are expected to be modified.  It is however unclear how
Alfv\'en, fast and slow waves modify the nonlinear processes that
govern plasma fluctuations in the presence of neutral dynamics. In the
context of cosmic ray propagation, Kulsrud \& Pearce (1969) noted that
the interaction of a neutral gas and plasma can damp Alfv\'en waves.
Neutrals interacting with plasma via a frictional drag process results
in ambipolar diffusion (Oishi \& Mac Low 2006). Ambipolar diffusion
plays a crucial role in the dynamical evolution of the near solar
atmosphere, interstellar medium, and molecular clouds and star
formation.  Oishi \& Mac Low (2006) found that ambipolar diffusion can
set a characteristic mass scale in molecular clouds. They found that
structures less than the ambipolar diffusion scale are present because
of the propagation of compressive slow mode MHD waves at smaller
scales. Leake et al (2005) showed that the lower chromosphere contains
neutral atoms, the existence of which greatly increases the efficiency
of wave damping due to collisional friction momentum transfer.  They
noted that Alfv\'en waves with frequencies above 0.6Hz are completely
damped and frequencies below 0.01 Hz are unaffected. They undertook a
quantitative comparative study of the efficiency of the role of
(ion-neutral) collisional friction, viscous and thermal conductivity
mechanisms in damping MHD waves in different parts of the solar
atmosphere.  They found that a correct description of MHD wave damping
requires the consideration of all energy dissipation mechanisms
through the inclusion of the appropriate terms in the generalized
Ohm’s law, the momentum, energy and induction equations.  Padoan et al
(2000) calculated frictional heating by ion-neutral (or ambipolar)
drift in turbulent magnetized molecular clouds and showed that the
ambipolar heating rate per unit volume depends on field strength for
constant rms Mach number of the flow, and on the Alfv\'enic Mach
number.

The role of ion-neutral collisions has been investigated extensively
by Balsara (1996) in the context of molecular clouds. Here the
momentum equation for the plasma is governed predominantly by the slow
comoving massive neutrals that tend to dissipate the Cloud's magnetic
field.  Because of the larger neutral to ion mass ratio, ion momentum
is entirely dominated by the neutral drag.  Hence the inertial terms
were ignored in plasma momentum equation. Balsara (1996) found that
slow waves propagate without significant damping on short wavelengths,
while the fast and Alfv\'en waves undergo rapid damping in
super-Alfv\'enic regimes.

In the heliospheric boundary regions, particularly the outer
heliosheath \cite[]{zank1999,dastgeer}, the plasma is mediated by
interstellar neutral gas by virtue of charge exchange. For example in
the local interstellar medium (LISM), the low density plasma and
neutral hydrogen (H) gas are coupled primarily through the process of
charge exchange.  In this process, both the total (proton+atom)
momentum and energy are conserved.  On sufficiently large temporal and
spatial scales, a partially ionized plasma is typically regarded as
equilibrated; this is the case for the LISM, but not for the outer
heliosheath, for example.

It is worth mentioning that the charge exchange process in its simpler
(and leading order) form can be treated like a friction or viscous
drag term in the fluid momentum equation, describing the relative
difference in the ion and neutral fluid velocities. The drag imparted
in this manner by a collision between ion and neutral also causes
ambipolar diffusion, a mechanism used to describe Alfv\'en wave
damping by cosmic rays (Kulsrud \& Pearce 1969) and also discussed by
Oishi \& Mac Low (2006) in the context of molecular clouds.

In this Letter, we focus on understanding the propagation
characteristics of MHD waves by considering collisional and charge
exchange interactions thus {\em including both simultaneously}.  Our
model is more general and fully nonlinear, compared to that of Balsara
(1996) which ignores nonlinear intertial terms in the plasma momentum
equation.  We retain this term and carry out fully time-dependent
simulations in two dimensions. We treat plasma and neutral fluid in a
self-consistent manner by employing the complete nonlinear
time-dependent fluid equations for both the fluids.  In Section 2, we
discuss the equations of a coupled plasma-neutral model, their
validity, the underlying assumptions and the normalizations.  Section
3 describes results of our nonlinear, coupled, self-consistent fluid
simulations. We find that the propagation characteristics of Alfv\'en
waves are altered signifincantly by the combined action of charge
exchange and collisional interaction processes. The latter leads to a
modulated growth and damping of the Alfv\'en and fast/slow modes.  In
Section 4, we summarize our results and describe physical processes
that may be responsible for the modulation process.

\section{Partially ionized Model Equations}
We assume that fluctuations in the plasma and neutral fluids are
isotropic, homogeneous, thermally equilibrated and turbulent. A
constant mean magnetic field is present. Local mean flows and
nonlinear structures may subsequently be generated by
self-consistently excited nonlinear instabilities. The boundary
conditions are periodic. This allows us to assume an infinite
partially ionized plasma where the length scale of characteristic
fluctuations ($\ell$) is much smaller than the size of plasma or
computational box ($L$) i.e. $\ell \ll L$. We further assume that
plasma particles interact with the neutral gas via collisions as well
as charge exchange. The intrinsic magnetized waves supported by the
plasma scatter plasma particles in a random manner thus sufficiently
isotropizing the plasma and neutral distribution functions
\cite[]{monin}. The latter enables us to use fluid descriptions for
both the plasma and neutral gas.  Most of these assumptions are
appropriate to realistic space and astrophysical turbulent flows. They
allow us to use MHD and hydrodynamic descriptions for the plasma and
the neutral components respectively.  In the context of the LISM and
the outer heliosheath plasmas, the plasma and neutral fluid remain
close to thermal equilibirium and behave as Maxwellian fluids.  Our
model simulates a plasma-neutral fluid that is coupled via collisions
and charge exchange in astrophysical plasmas.  The fluid model
describing nonlinear turbulent processes, in the presence of charge
exchange and collision, can be described in terms of the plasma
density ($\rho_p$), velocity (${\bf U}_p$), magnetic field (${\bf
  B}$), pressure ($P_p$) components according to the conservative form

\be
\label{mhd}
 \frac{\partial {\bf F}_p}{\partial t} + \nabla \cdot {\bf Q}_p={\cal Q}_{p,n},
\ee
where,
\[{\bf F}_p=
\left[ 
\begin{array}{c}
\rho_p  \\
\rho_p {\bf U}_p  \\
{\bf B} \\
e_p
  \end{array}
\right], 
{\bf Q}_p=
\left[ 
\begin{array}{c}
\rho_p {\bf U}_p  \\
\rho_p {\bf U}_p {\bf U}_p+ \frac{P_p}{\gamma-1}+\frac{B^2}{8\pi}-{\bf B}{\bf B} \\
{\bf U}_p{\bf B} -{\bf B}{\bf U}_p\\
e_p{\bf U}_p
-{\bf B}({\bf U}_p \cdot {\bf B})
  \end{array}
\right],\]
\[ {\cal Q}_{p,n}=
\left[ 
\begin{array}{c}
0  \\
{\bf Q}_{M,p,n} + {\bf{F}}_{p,n}   \\
0 \\
Q_{E,p,n} + {\bf U}_p \cdot {\bf{F}}_{p,n}
  \end{array}
\right]
\] 
and
\[ e_p=\frac{1}{2}\rho_p U_p^2 + \frac{P_p}{\gamma-1}+\frac{B^2}{8\pi}.\]
Note the presence of the source terms $Q$ that couple the plasma
self-consistently to the neutral gas.  The above set of plasma
equations is supplimented by $\nabla \cdot {\bf B}=0$ and is coupled
self-consistently to the neutral density ($\rho_n$), velocity (${\bf
  V}_n$) and pressure ($P_n$) through a set of hydrodynamic fluid
equations,
\be
\label{hd}
 \frac{\partial {\bf F}_n}{\partial t} + \nabla \cdot {\bf Q}_n={\cal Q}_{n,p},
\ee
where,
\[{\bf F}_n=
\left[ 
\begin{array}{c}
\rho_n  \\
\rho_n {\bf V}_n  \\
e_n
  \end{array}
\right], 
{\bf Q}_n=
\left[ 
\begin{array}{c}
\rho_n {\bf V}_n  \\
\rho_n {\bf V}_n {\bf V}_n+ \frac{P_n}{\gamma-1} \\
e_n{\bf V}_n
  \end{array}
\right],\]
\[{\cal Q}_{n,p}=
\left[ 
\begin{array}{c}
0  \\
{\bf Q}_{M,n,p} + {\bf{F}}_{n,p}   \\
Q_{E,n,p} + {\bf V}_n \cdot {\bf{F}}_{n,p}
  \end{array}
\right],
\] 
\[e_n= \frac{1}{2}\rho_n V_n^2 + \frac{P_n}{\gamma-1}.\]
Equations (\ref{mhd}) to (\ref{hd}) form an entirely self-consistent
description of the coupled  plasma-neutral turbulent fluid in a
partially ionized medium.

Several points are worth noting.  The charge-exchange momentum sources
in the plasma and the neutral fluids, i.e.  \eqs{mhd}{hd}, are
described respectively by terms ${\bf Q}_{M,p,n}({\bf U}_p,{\bf V}_n,\rho_p,
\rho_n, T_n, T_p)$ and ${\bf Q}_{M,n,p}({\bf V}_n,{\bf U}_p,\rho_p, \rho_n,
T_n, T_p)$. These expressions, described in \cite{pauls1995,dastgeer},
have the following form.
\be
\label{qm2}
 {\bf Q}_{M,p,n}({\bf V}_n,{\bf U}_p) = m\sigma n_{p}n_n ({\bf V}_n -{\bf
   U}_p) \left[ U^\ast + \frac{V_{T_n}^2}{\delta V_{{\bf U}_p,{\bf
         V}_n}} -\frac{V_{T_p}^2}{\delta V_{{\bf V}_n,{\bf U}_p}}
   \right],
\ee 
\eqa
\label{qe}
Q_{E,p,n}({\bf V}_n,{\bf U}_p) = \frac{1}{2}m \sigma  n_{p}n_n U^\ast (V_n^2-U_p^2)+
\frac{3}{4}m  \sigma  n_{p}n_n (V_{T_{n}}^2\Delta V_{{\bf U}_p,{\bf V}_n}-V_{T_{p}}^2
\Delta V_{{\bf V}_n,{\bf U}_p}) 
 \nonumber \\
-m \sigma  n_{p}n_n
\left[ {\bf V}_n \cdot ({\bf U}_p-{\bf V}_n)\frac{V_{T_n}^2}{\delta V_{{\bf U}_p,{\bf V}_n}}-
{\bf U}_p \cdot ({\bf V}_n-{\bf U}_p)\frac{V_{T_{p}^2}}{\delta V_{{\bf V}_n,{\bf U}_p}}\right],
\eeq
with
\[U^\ast=U_{{\bf
  U}_p,{\bf V}_n}^\ast = U_{{\bf V}_n,{\bf U}_p}^\ast =
  \left[\frac{4}{\pi}V_{T_{p}}^2+\frac{4}{\pi}V_{T_{n}}^2 +\Delta
  U^2\right]. \]
 A swapping of the plasma and the neutral fluid
velocities in this representation corresponds, for instance, to
momentum changes (i.e. gain or loss) in the plasma fluid as a result
of charge exchange with the neutral atoms (i.e. ${\bf Q}_{M,p,n}({\bf
  U}_p,{\bf V}_n,\rho_p, \rho_n, T_n, T_p)$ in
Eq. (\ref{mhd})). Similarly, momentum change in the neutral fluid by
virtue of charge exchange with the plasma ions is described by ${\bf
  Q}_{M,n,p}({\bf V}_n,{\bf U}_p,\rho_p, \rho_n, T_n, T_p)$ in
Eq. (\ref{hd}).

The plasma-neutral collisional forces are modeled by employing the
following forcing term in \eqs{mhd}{hd}.
\be
\label{col}
{\bf{F}}_{p,n} = \nu \rho_p \rho_n ({\bf U}_p-{\bf
  V}_n) = - {\bf{F}}_{n,p},
\ee
where $\nu$ is the collision frequency.  The collisions modify plasma
and neutral momentum and energy, but not the density.  The negative
sign before ${\bf{F}}_{n,p}$ in \eq{col} ensures conservation of total
momentum.  We derive modified equations of energy for both fluids
following the treatment described in \cite{ll}.  The collisional force
is self-consistently calculated in our model. Note also the charge
exchange source term in momentum and energy of both the fluids.  In
the absence of collision and charge exchange interactions, the plasma
and the neutral fluid are de-coupled trivially and behave as ideal
fluids.  Like collisions, charge-exchange interactions modify the
momentum and the energy of plasma and the neutral fluids and conserve
density in both the fluids (since we neglect photoionization and
recombination). Nonetheless, the volume integrated energy and the
density of the entire coupled system will remain conserved in a
statistical manner. The conservation processes can however be altered
in the presence of any external forces. These can include large-scale
random driving of turbulence due to external forces such as supernova
explosions, stellar winds, etc and instabilities. Finally, the
magnetic field evolution is governed by the usual induction equation,
i.e. Eq. (\ref{mhd}), that obeys the frozen-in-field theorem unless
some nonlinear dissipative mechanism introduces small-scale damping.

The underlying coupled fluid model can be non-dimensionalized
straightforwardly using a typical scale-length ($\ell_0$), density
($\rho_0$) and velocity ($v_0$). The normalized plasma density,
velocity, energy and the magnetic field are respectively;
$\bar{\rho}_p = \rho_p/\rho_0, \bar{\bf U}_p={\bf U}_p/v_0,
\bar{P}_p=P_p/\rho_0v_0^2, \bar{\bf B}={\bf B}/v_0\sqrt{\rho_0}$. The
corresponding neutral fluid quantities are $\bar{\rho}_n =
\rho_n/\rho_0, \bar{\bf U}_n={\bf U}_n/v_0,
\bar{P}_n=P_n/\rho_0v_0^2$. The momentum and the energy
charge-exchange terms, in the normalized form, are respectively
$\bar{\bf Q}_m={\bf Q}_m \ell_0/\rho_0v_0^2, \bar{Q}_e=Q_e
\ell_0/\rho_0v_0^3$. The non-dimensional temporal and spatial
length-scales are $\bar{t}=tv_0/\ell_0, \bar{\bf x}={\bf
  x}/\ell_0$. Note that we have removed bars from the set of
normalized coupled model equations (\ref{mhd}) \& (\ref{hd}).  The
charge-exchange cross-section parameter ($\sigma$), which does not
appear directly in the above set of equations, is normalized as
$\bar{\sigma}=n_0 \ell_0 \sigma$, where the factor $n_0\ell_0$ has
dimension of (area)$^{-1}$.  By defining $n_0, \ell_0$ through
$\sigma_{ce}=1/n_0\ell_0=k_{ce}^2$, we see that there exists a charge
exchange mode ($k_{ce}$) associated with the coupled plasma-neutral
turbulent system.  For a characteristic density, this corresponds
physically to an area defined by the charge exchange mode being equal
to (mpf)$^2$.  Thus the larger the area, the higher is the probability
of charge exchange between plasma ions and neutral atoms.  Therefore,
the probability that charge exchange can directly modify those modes
satisfying $k<k_{ce}$ is high compared to modes satisfying $k>k_{ce}$.
Since the charge exchange length-scales are much smaller than the
turbulent correlation scales, this further allows many turbulent
interactions amongst the nonlinear turbulent modes before they undergo
at least one charge exchange.  An exact quantitive form of sources due
to charge exchange in our model is taken from Shaikh \& Zank (2008).

\section{Evolution of MHD waves}
We now investigate the nonlinear evolution of the coupled
plasma-neutral system described by \eqs{mhd}{hd}. Our goal is to
understand the evolution of MHD waves in a partially ionized turbulent
plasma, mediated by the complex coupling between the two
distinguishable fluids. For this purpose, we initialize the field
variables (i.e. velocity, magnetic, density, energy and pressure
fields) in the plasma and neutral fluids with a random, uncorrelated,
out-of-phase distribution in space. The initial ratio of the plasma
and neutral density is unity.  The plasma and neutral fluids
initialized in this manner characterize spatially local turbulence in
the partially ionized astrophysical plasma.  These random fields
develop spatially and temporally according to the set of \eqs{mhd}{hd}
and forces due to charge exchange and collision dictate nonlinear
interaction processes. In ideal MHD and hydrodynamic fluids
(i.e. without the coupling forces in \eqs{mhd}{hd}), known quadratic
conserved quantities like energy, helicity, vorticity etc
predominantly govern the spectral transfer of energy across disparate
characteristic scales in the inertial range. Their role is well
studied within the paradigm of statistical theories of turbulence
\cite[]{kol,iros,krai,monin,biskamp03}. By contrast, the complex
nature of the coupling forces in \eqs{mhd}{hd} poses a formidable
hurdle to deriving quadratic conserved quantities.  It is not clear to
us as how these quantities will influence spectral transfer in the
inertial range. Furthermore, the driving forces in \eqs{mhd}{hd} may
presumably be operative at any length and time scales. It is unclear
how this will alter the conventional dissipative and diffusive
processes in a partially ionized plasma. Unlike the constant drag
force (i.e. not changing in space and time) used before for a neutral
fluid; and which acts to damp Alfv\'en waves in the partially ionized
plasma (Kulsrud \& Pearce 1969, Balsara 1996, Shaikh 2010), it is not
clear how the complex, nonlinear, time and space dependent driving
forces in \eqs{mhd}{hd} will modify the propagation and interactions
characteristic of Alfv\'en, fast and slow MHD waves or even
compressive sound waves in the coupled neutral gas.

To address the above issues, we have developed a two-dimensional (2D)
nonlinear fluid code to numerically integrate \eqs{mhd}{hd}. The 2D
simulations are not only computationally less expensive (compared to a
fully 3D calculation), but they offer significantly higher resolution
(to compute inertial range turbulence spectra) even on
moderately-sized clusters like our local Beowulf system.  The spatial
discretization in our code uses a discrete Fourier representation of
turbulent fluctuations based on a spectral method, while we use a
4th-order Runge-Kutta method for the temporal integration. All the
fluctuations are initialized isotropically with random phases and
amplitudes in Fourier space.  A mean constant magnetic field $B_0$ is
assumed along the $y$-direction.  Our algorithm ensures conservation
of total energy and mean fluid density per unit time in the absence of
charge exchange and external random forcing. Additionally, $\nabla
\cdot {\bf B}=0$ is satisfied at each time step.  Our code is
massively parallelized using Message Passing Interface (MPI) libraries
to facilitate higher resolution. The initial isotropic turbulent
spectrum of fluctuations is chosen to be close to $k^{-2}$ with random
phases in all three directions.  The choice of such a (or even a
flatter than -2) spectrum does not influence the dynamical evolution
as the final state in our simulations progresses towards fully
developed turbulence.

\begin{figure}[t]
\includegraphics[width=8cm]{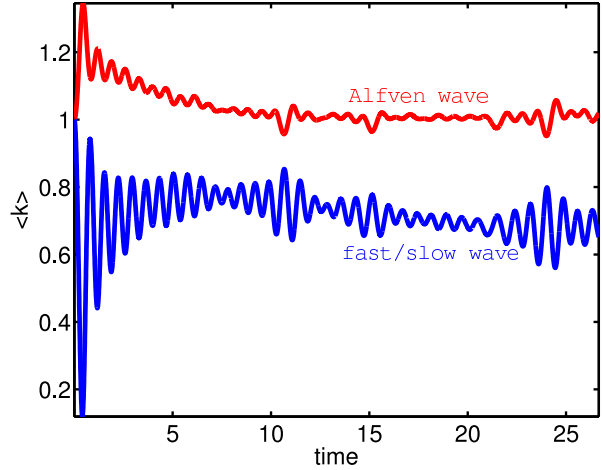}
\caption{ Evolution of modal energy in
  Alfv\'en and fast/slow waves is shown by their respective average
  mode behavior. The oscillatory behavior of the average mode
  associated with Alfv\'en and fast/slow indicates that charge
  exchange and collisions with neutrals damp the Alfv\'en and
  fast/slow wave which are then reexcited by an instability
  (growth). This process repeats itself during the entire
  evolution.}
\label{fig1}
\end{figure}

\begin{figure}[t]
\includegraphics[width=8cm]{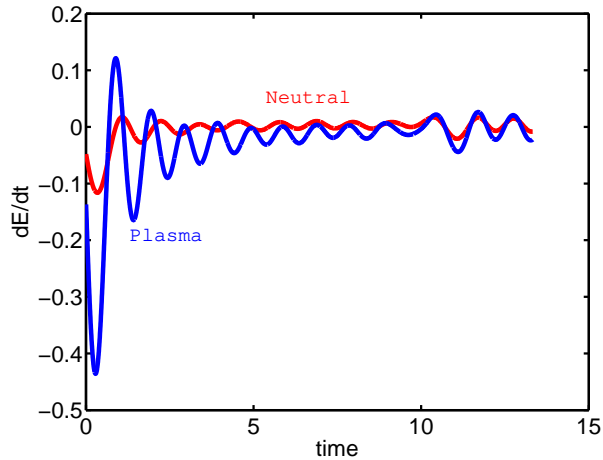}
\caption{ Energy exchange rate between the plasma
  and neutral fluids shows that the rate of transfer of energy between
  the two fluids is constant. We plot the corresponding energy equations
from Eqs (1) \& (2).}
\label{fig2}
\end{figure}

From our time-dependent nonlinear fluid simulations, we determine the
quantative evolution of Alfv\'en and fast/slow waves in a partially
ionized environment. For this purpose, we distinguish Alfv\'enic and
non-Alfv\'enic, i.e.  corresponding to the compressional or due to
slow and fast magnetosonic modes, contributions to the turbulent
velocity fluctuations. Thus, we introduce diagnostics that distinguish
between the modes corresponding to Alfv\'enic and slow/fast
magnetosonic fluctuations.  Since the Alfv\'enic fluctuations are
transverse \cite[]{biskamp03,dastgeer}, the propagation wave vector is
orthogonal to the velocity field fluctuations i.e.  ${\bf k} \perp
{\bf U}$, and the average spectral energy contained in these (shear
Alfv\'enic modes) fluctuations can be computed as \cite[]{dastgeer2}
\be
\label{alf}
 \langle k_{A} (t) \rangle \simeq \sqrt{\frac{{\sum_{\bf k}}|i {\bf k} 
\times {\bf U}_{\bf k}|^2}{\sum_{\bf k} |{{\bf U}_{\bf k}}|^2}}.
\ee
The above relationship yields a finite spectral contribution from the
$| {\bf k} \times {\bf U}_{\bf k}|$ characteristic turbulent
Alfv\'enic modes.  It is essentially a modal energy associated with  the
Alfv\'enic fluctuations that obey \eq{alf} in the inertial range
turbulent spectrum.  On the other hand,
fast/slow (i.e. compressive) magnetosonic modes propagate
longitudinally along the velocity field fluctuations, i.e.  ${\bf k}
\parallel {\bf U}$ and thus carry a finite component of energy
corresponding {\em only} to the $i {\bf k} \cdot {\bf U}_{\bf k}$ part
of the velocity field, which can be determined from the following
relationship \be
\label{fs}
 \langle k_{c} (t) \rangle \simeq \sqrt{\frac{{\sum_{\bf k}}|i {\bf k} 
\cdot {\bf U}_{\bf k}|^2}{\sum_{\bf k} |{{\bf U}_{\bf k}}|^2}}.
\ee
The expression for $k_{c}$ essentially describes the modal energy
contained in the non-solenoidal component of the MHD turbulent modes.

The evolution of an average mode associated with Alfv\'enic and
compressive waves is shown in \Fig{fig1}.  We vary charge exchange
strength and collision parameter in our simulation to examine their
effects on the propagation of Alfv\'enic and compressive modes in a
partially ionized plasma. Clearly, the two processes operate on
different time and length scales and are self-consistently modeled in
our simulations.  We find that charge exchange and collisional
interactions jointly {\em modulate} Alfv\'en and fast/slow modes. This
is shown in \Fig{fig1} for $512^2$ modes in a two dimensional box of
length $2\pi \times 2\pi$. The other parameters in our simulations are
charge exchange wavenumber $k/k_{ce} \sim 0.01$, fixed time step
$dt=10^{-3}$, and collision parameter $\nu \sim 0.001$. The background
normalized constant magnetic field $B_0=0.5$. Our simulations are
fully nonlinear and the ratio of the mean and fluctuating magnetic
fields $\delta {\bf B}/{\bf B}_0 \sim 1$.

The modulation process depicted in \Fig{fig1} can be understood on the
basis of charge exchange and collisional interactions, together with
convective nonlinearities. During the modulation, there exists a rapid
onset of the average mode associated with Alfv\'enic waves. This onset
is triggered essentially by linear instability process. Our linear
stability analysis (not described here) indicates that the underlying
coupled plasma-neutral system possesses several unstable modes. These
modes account for the growing as well as damping of Alfv\'enic and
fast/slow compressive waves. As illustrated in \Fig{fig1}, a
concurrent damping of the fast/slow compressive mode preceding the
growth of Alfv\'enic mode occurs.  When the two modes reach their
extremal rise or fall, they reverse their behavior. The fast/slow
compressive mode begins to rise at the expense of damping of
Alfv\'enic mode. This process continues non periodically and repeats
itself in time thereby exhibiting a nearly oscillatory behavior.  It
follows from \Fig{fig1} that the two modes, i.e. Alfv\'enic and
fast/slow compressive, regulate each other in a predator-prey
manner. Dynamically, the Alfv\'enic mode grows at the expense of
fast/slow mode. When Alfv\'enic mode reaches its maximum amplitude,
where the fast/slow mode remains at its lowest magnitude, the latter
eats up the Alfv\'enic mode and vice versa.  This process is
repetitive. It is also noteworthy that the overall amplitude of both
the modes decays in time and regrows later.

The growth and damping of wave energy can further be understood from
the energy transfer rates ($dE/dt$) between the plasma and neutral
fluids. This is shown in \Fig{fig2}. It is evident from the figure
that the total energy of the plasma increases at the expense of the
neutral. When the waves damp, the total energy of the plasma decreases
while it increases for the neutrals. The energy transfer rates also
exhibit a periodic behavior following the modulation of energy in
Alfv\'en and fast/slow modes.  Nonetheless, the energy transfer rates
try to approach nearly zero i.e. $dE/dt \rightarrow 0$, but are not
entirely zero due to the modulation process. This is an interesting
result that differs from pure (i.e. uncoupled) plasma or neutral fluid
turbulence \cite[]{monin,biskamp03}. The latter exhibits $dE/dt \simeq
0$ thereby leading to a constant transfer of energy across inertial
range fluctuations. This further demonstrates that the total energy of
the neutral and plasma fluid is nearly constant during the steady
state (See Fig 2).  What we find in our simulations is a
modulated-energy-transfer process that is completely different from
the constant energy transfer rate. The energy transfer rates in
\Fig{fig2} thus provide a self-consistent description of the
modulation process observed in our simulations.

\section{Conclusion}
In summary, we developed a two dimensional self-consistent model of
plasma and neutral fluids that is coupled through charge exchange and
collisional interactions.  Notably, we find that these interactions
not only modify the convective nonlinear interactions in the coupled
\eqs{mhd}{hd}, but they also influence the propagation characteristics
of MHD waves significantly.  To determine the evolution of
characteristic MHD waves, we distinguish the Alfv\'enic and
non-Alfv\'enic contributions of MHD modes in the partially ionized
turbulent environment. It is found that charge exchange and
collisional interactions lead to a modulation of Alfv\'enic and
fast/slow compressive modes. This is unlike purely collisional
processes described by \cite{kulsrud} and \cite{balsara} that tend to
damp Alfv\'enic modes. By contrast, inclusion of the charge exchange
interactions leads to an entirely different scenario and it alters the
propagation characteristics of MHD waves. Charge exchange, in the
presence of collisions and convective nonlinear interactions, leads to
an alternate growth and damping (i.e. modulation) of Alfv\'enic and
fast/slow compressive modes.  It is further noted that the modulation
of waves is mediated by the conservation of momentum because of the
characteristic wave velocity that is involved in \eqs{alf}{fs}. It is
the momentum transfer function ($Q_m$) that is responsible for
regulating the rise/fall of the amplitude of the two waves and
maintains the conservation process. This point is further consistent
with Fig (1). We finally comment on the conservation of momentum and
energy in our simulations. It seems from Fig (2) that the term $dE/dt$
hovers around zero, but does not fully converge to zero. We learn that
due to the small scale self-consistent dissipation (or numerical
dissipation), it is difficult to observe a 100\% conservation. We
continue to work to improve this conservation in our model.

Our results should find application to a variety of astrophysical
environments in which a partially ionized plasma is typical. Examples
include the outer heliosheath formed by the interactions of the solar
wind with the local interstellar medium, the magnetic collapse of
molecular clouds and star formation and the general transfer of energy
in partially ionized plasmas surrounding other stellar systems
\cite[]{wood}.

\section*{Acknowledgments}
We acknowledge the partial support of NASA grants NNX09AB40G,
NNX07AH18G, NNG05EC85C, NNX09AG63G, NNX08AJ21G, NNX09AB24G,
NNX09AG29G, and NNX09AG62G.

\end{document}